# Quasiparticle Dynamics in Reshaped Helical Dirac Cone of Topological Insulators

Lin Miao[a,1], Z. F. Wang[b,1], Wenmei Ming[b], Meng-Yu Yao[a], Mei-Xiao Wang[a], Fang Yang[a], Y. R. Song[a], Fengfeng Zhu[a], Alexei V. Fedorov[c], Z. Sun[d], C. L. Gao[a], Canhua Liu[a], Qi-Kun Xue[e], Chao-Xing Liu[f], Feng Liu[b,2], Dong Qian[a,2], Jin-Feng Jia[a,2]

[a] Key Laboratory of Artificial Structures and Quantum Control (Ministry of Education), Department of Physics, Shanghai Jiao Tong University, Shanghai 200240, China

[b] Department of Materials Science &Engineering, University of Utah, Salt Lake City, Utah 84112, USA

[c] Advanced Light Source, Lawrence Berkeley National Laboratory, Berkeley, California 94305, USA

[d] National Synchrotron Radiation Laboratory, University of Science and Technology of China, Hefei, 230026, China

[e] State Key Laboratory for Low-Dimensional Quantum Physics, Department of Physics, Tsinghua University, Beijing 100084, China

[f] Department of Physics, Penn State University, University Park, PA 16802, USA

[1] These authors contributed equally to this work

[2] To whom correspondence should be addressed. Email: dqian@sjtu.edu.cn, fliu@eng.utah.edu, jfjia@sjtu.edu.cn

# Abstract

Topological insulators (TIs) and graphene present two unique classes of materials which are characterized by spin polarized (helical) and non-polarized Dirac-cone band structures, respectively. The importance of many-body interactions that renormalize the linear bands near Dirac point in graphene has been well recognized and attracted much recent attention. However, renormalization of the helical Dirac point has not been observed in TIs. Here, we report the experimental observation of the renormalized quasi-particle spectrum with a skewed Dirac cone in a single Bi bilayer grown on $Bi_2Te_3$ substrate, from angle-resolved photoemission spectroscopy. First-principles band calculations indicate that the quasi-particle spectra are likely associated with the hybridization between the extrinsic substrate-induced Dirac states of Bi bilayer and the intrinsic surface Dirac states of $Bi_2Te_3$ film at close energy proximity. Without such hybridization, only single-particle Dirac spectra are observed in a single Bi bilayer grown on $Bi_2Se_3$, where the extrinsic Dirac states Bi bilayer and the intrinsic Dirac states of $Bi_2Se_3$ are well separated in energy. The possible origins of many-body interactions are discussed. Our findings provide a means to manipulate topological surface states.

**Introduction**

Much recent attention has been devoted to graphene (1-6) and TIs (7-18), two unique material systems that exhibit conical linear electron bands of Dirac spectra. Quasi-particles of Dirac Fermions are distinct from those of ordinary Fermi liquids (19-21). Although rather difficult and rare, recent angle-resolved photoemission spectroscopy (ARPES) experiments (1,2,5,6) have directly shown the existence of many-body quasi-particle spectra near Dirac point in graphene, manifesting electron-electron, electron-phonon and electron-plasmon interactions. Similar to graphene, TIs also possess Dirac cone, albeit it is spin-polarized or helical Dirac cone. So far, however, no renormalized quasi-particle spectra near the helical Dirac point similar to graphene have been reported in any known TIs, and most studies of TIs are based on the single-particle picture (9,11,12,14,18). Here, for the first time, we report direct experimental observation of a skewed helical Dirac point, a signature quasi-particle spectrum indicative of many-body interactions, by ARPES in a novel TI system of Bi (111) bilayer grown on $Bi_2Te_3$ substrate, where a 2D TI is interfaced with a 3D TI (22).

ARPES can probe the quasi-particle's scattering rate at different energy scales, and therefore can access the many-body interactions directly (23). Our experimental observation of the quasi-particle spectra manifesting many-body effects is characterized with a "vertically non-dispersive" feature near Dirac point. Based on model DFT calculations of electron bands as a function of the artificially changed interfacial distance between the Bi bilayer and substrate, we found that the renormalized quasi-particle spectra in $Bi/Bi_2Te_3$ is likely associated with the strong hybridization between the substrate-induced Dirac states of Bi bilayer and the surface Dirac states of $Bi_2Te_3$ substrate at close energy proximity. When these two Dirac states are well separated in energy without the hybridization, such as in $Bi/Bi_2Se_3$, only single-particle Dirac spectra are observed without the feature of many-body interactions. We further discuss possible physical origins of the observed many-body spectra, and we are able to exclude the electron-phonon interaction.

## Results and Discussion

We have grown Bi (111) films in the layer-by-layer mode on the (111)-oriented $Bi_2Te_3$ and $Bi_2Se_3$ substrates. The growth mode is studied by reflection high-energy electron diffraction (RHEED) and scanning tunneling microscopy (STM) (details can be found in the Supplementary Information). The in-plane lattice constant is measured to match the substrate exactly with a perfect coherent interface, so that the Bi (111) film is under 3.5% and 9.0% tensile strain on $Bi_2Te_3$ or $Bi_2Se_3$, respectively. Here, we focus on electronic properties of single Bi (111) bilayer grown on both substrates by measuring the electron band structures using ARPES.

The band structures of the bare $Bi_2Te_3$ and $Bi_2Se_3$ substrates near the Fermi level around zone center (Γ point) are presented in Fig. 1*A* and 1*F*, respectively. Linearly dispersive energy bands from surface states forming a Dirac cone at the Γ point (14) are well separated from bulk bands. Known from previous ARPES studies and first-principles calculations (13,14), the Dirac point of $Bi_2Te_3$ is hidden by the "M" shape bulk valence bands. In our films, the hidden Dirac point can be located at ~ 0.2 eV below Fermi level by extrapolating the "V" shape surface bands. Similar to previous study (12), the Dirac point of $Bi_2Se_3$ is inside the bulk gap, located at ~ 0.3 eV below Fermi level.

The electronic band structures are observed to change dramatically when a single Bi (111) bilayer is grown. On $Bi_2Te_3$, the "M" shape bulk bands disappear, as seen in Fig. 1*B* and 1*C*. Most surprisingly, we see two sets of linearly dispersive bands crossing at Γ point, one at the energy slightly above 0.2 eV and the other slightly below 0.2 eV, which is the location of the hidden Dirac point of bare $Bi_2Te_3$ film (Fig. 1*A*). In a recent study, it was thought there was only one Dirac point in this energy range (22) due to the limited ARPES momentum resolution. With substantial improvement of film quality and ARPES resolution, two Dirac crossing points are clearly resolved, with a magnified view shown in Fig. 1*D*. The upper "V" and lower "Λ" Dirac cone do not touch, and most interestingly, there appears a vertically non-dispersive feature between them with an energy width of ~ 0.05 eV. This can also be clearly seen in the momentum distribution curves (MDCs), as shown in Fig. 1$E$. In this "vertical" region, MDCs only have a single peak. The Fermi velocity ($v_F$) of the "V" and "Λ" bands is ~ 3.2×10$^5$ m/s and ~ 4.5×10$^5$ m/s, respectively. The $v_F$ of "V" band in the bare Bi$_2$Te$_3$ film (Fig. 1$A$) is ~ 4.5 × 10$^5$ m/s. Electrons between "V" and "Λ" bands would have infinite velocity if the measured signals are real single-particle spectra. We do not believe that is the case.

The renormalization of linear dispersive Dirac cone is a well-known signature of quasi-particle spectrum arising from many-body interactions (1-6). We found that such quasi-particle TI spectrum is very unique to the Bi/Bi$_2$Te$_3$ system. Even in a similar and closely related Bi/Bi$_2$Se$_3$ system, only ordinary single-particle TI spectra are observed as shown in Figs. 1$G$ and 1$H$. Two well separated Dirac points at different energies (marked as D$_S$ and D$_{Bi}$) are seen in Fig. 1$H$ and 1$G$. D$_S$ is located at ~0.3 eV below Fermi level, almost at the same position as the Dirac point in the bare Bi$_2$Se$_3$ (Fig. 1$F$). $v_F$ near D$_S$ is ~2.8×10$^5$ m/s which is about half of $v_F$ in bare Bi$_2$Se$_3$ film (5.7×10$^5$ m/s, Fig. 1$F$). D$_{Bi}$ is very close to Fermi level with a $v_F$~ 5.3 × 10$^5$ m/s. Most noticeably, no vertically non-dispersive feature is visible in either D$_S$ or D$_{Bi}$.

The 2D character of the observed linearly dispersive bands is further confirmed by the photon-energy-dependent experiments that are widely used to separate surfaces states from bulk bands (11). By tuning incident photon energy, we change the detectable momentum along the film's normal direction ($k_z$). Consequently, the measured energy band dispersions from bulk contributions will change as a function of incident photon energy. Fig. 2 shows the ARPES spectra under different incident photon energies (more spectra can be found in the Supplementary Information). Except for the relative spectral weight or intensity, the band dispersions and the location of Dirac cone do not change. This indicates that the observed energy bands are coming only from 2D Bi bilayer and/or surface states of TI substrates.

The observation of quasi-particle TI spectrum and the fact that it only occurs in Bi/Bi$_2$Te$_3$ but not in Bi/Bi$_2$Se$_3$ are both very intriguing. To help understand the physical origin of the quasi-particle TI spectrum in Bi/Bi$_2$Te$_3$ and the underlying difference between these two systems, we have performed DFT calculations of electron band structures of both systems. Figs. 3*A-D* show the calculated bands of Bi/Bi$_2$Te$_3$ and Bi/Bi$_2$Se$_3$ systems along high symmetry directions. In both systems, the Dirac cone structures are all helical Dirac cones (see Supplementary Information). Fig. 3*E* and 3*F* show the experimental bands overlaid with the calculated bands. We see that overall the agreement of band energies and dispersions between experiment and calculation is very good except the slightly shifted Fermi level. However, there is one significant discrepancy in Fig. 3*E*: the experiment shows a non-dispersive feature between two vertically separated "V" and "Λ" Dirac cones as discussed above, while the calculation shows two Dirac cones crossing at one point as for a typical single-particle Dirac-cone spectrum. This indicates that *the observed non-dispersive feature is originated from many-body interactions that cannot be reproduced by DFT calculations of single-particle spectrum.*

To further support the above point, we have extracted the self-energies from the experimental data and incorporate them into the calculated single-particle spectral functions (see magnified view near Γ point in Fig. 3*G*) to construct the quasi-particle spectra (details can be found in the Supplementary Information). Fig. 3*H* shows the resulting quasi-particle spectral function with the self-energy correction near Γ point. The single-point-crossing Dirac cone (Fig. 3*G*) elongates vertically into a non-dispersive feature between two Dirac "V" and "Λ" points (Fig. 3*H*), in good agreement with the experimental spectra, as shown in Fig. 3*I*, where the experimental bands are overlaid with the theoretical bands. Here, the purpose of our theoretical fitting is mainly to show the qualitative importance of the self-energy correction, but the extracted self-energies should be treated with caution. This is because accurate quantitative values will need to be extracted from much more extended higher-resolution ARPES data. On the other hand, as seen from Fig. 3*C* and 3*D*, there are two Dirac cones in Bi/Bi$_2$Se$_3$ system.

We further performed spectral analysis to better understand the DFT band structures. The projected spectral function calculations show that the calculated Dirac cone at ~ 0.15 eV below Fermi level (Fig. 3*A* and 3*B*) is a hybrid Dirac state between the Bi bilayer and the bulk Bi$_2$Te$_3$ film, with ~50% spectral weight coming from the Bi bilayer. This is very surprising, considering the fact that the single Bi (111) bilayer is well-known to have a finite gap (24-27). It turned out to be caused by a hybridization of two Dirac states, one intrinsic from Bi$_2$Te$_3$ substrate and the other extrinsic from Bi bilayer induced by the interface (see discussion below). A recent study had assigned this Dirac cone to Bi$_2$Te$_3$ without knowing the second Bi Dirac cone, although they did see the charge density of this Dirac state leaked into Bi (22). In Bi/Bi$_2$Se$_3$ system, 90% of its spectral weight of the Dirac cone at ~ 0.1 eV above the Fermi level (Fig. 3*C* and 3*B*) comes from the Bi bilayer with little hybridization with the substrate states. The second Dirac cone at ~ 0.2 eV below the Fermi level fully comes from Bi$_2$Se$_3$.

If one looks at the calculated band structure of Bi/Bi$_2$Te$_3$ (Figs. 3A and 3B) alone, there appeared only one Dirac point, presumably coming from the Bi$_2$Te$_3$ substrate. In contrast, the calculated band structure of Bi/ Bi$_2$Se$_3$ (Figs. 3C and 3D) clearly shows two Dirac points. To resolve this difference, we have performed a set of "model" calculations by artificially increasing the interfacial distance between the Bi bilayer and Bi$_2$Te$_3$ (or Bi$_2$Se$_3$) substrate to gradually tune the interface coupling strength, as shown in Fig. 4 and Fig. 5. These systematic model calculations reveal that both systems have a second Dirac point produced by the Bi bilayer due to interfacial interaction. Interestingly, it just happened that at the equilibrium distance, the interface induced Bi Dirac point lies at almost the same energy as the Bi$_2$Te$_3$ Dirac point, so it appeared as if there was only one Dirac point in Fig. 4E, as we discuss below.

The bands in Fig. 4D at large interface separation represent essentially the strained free-standing Bi bilayer bands at the $Bi_2Te_3$ lattice constant. Projecting the spectral functions onto the top Bi bilayer with the decreasing interfacial distance (from Fig. 4D to Fig. 4A), we found that the interaction between the Bi and substrate gradually splits the degenerated Bi bands (Fig. 4D), and the Dirac cone ($D_h$ in Fig. 4A) partially comes from the lower branch of the Bi band. We also projected the spectral functions onto the top Bi bilayer plus upper 2QL $Bi_2Te_3$ (Fig.4 E-H), to see what happens to the "bulk" Dirac cone ($D_S$) of $Bi_2Te_3$ film in this process. In Fig. 4H (4Å away from the equilibrium distance), we can clearly see the bulk Dirac cone. But with the decreasing interfacial distance (from Fig. 4H to Fig. 4E), the interaction blurs the bulk Dirac cone (Fig. 4H) making it indistinguishable from the substrate-induced Bi bilayer Dirac cone at the equilibrium distance (Fig. 4E), and the two hybridize into the $D_h$.

For comparison, the $Bi/Bi_2Se_3$ results are shown in Fig. 5. The bands in Fig. 5D at large interface separation represent essentially the strained free-standing Bi bilayer bands at the $Bi_2Se_3$ lattice constant, which is to be noted different from Fig. 4D. Similar to the $Bi/Bi_2Te_3$ system, there is also a substrate-induced Dirac cone ($D_{Bi}$) forming from the lower branch of the Bi bilayer band; but different from the $Bi/Bi_2Te_3$ system, its position is about ~ 0.1 eV above the Fermi level (Fig. 5A) and about 90% of its spectral weight comes from the Bi bilayer having little hybridization with the substrate states. Changing the interfacial distance, the position of the bulk $Bi_2Se_3$ Dirac cone ($D_S$) is almost unchanged staying at ~ 0.2 eV below the Fermi level, as shown in Fig. 5 E-H; its spectral weight remains ~100% from $Bi_2Se_3$ independent of interfacial spacing. This is consistent with the ARPES experiment observing two Dirac cones at about these two energies (Figs. 1G and 1H).

From the dependence of the spectral functions on the interfacial distance in Fig. 4 and 5, we find an interesting point that the original bulk $Bi_2Te_3$ Dirac cone and the newly induced Bi Dirac cone coincidentally lie together at close energy proximity (both at ~ 0.2 eV below the Fermi level). Since the states around the two Dirac cones are all the surface states (2D states, Fig. 2), we suggest that the energy resonance between them leads to the enhanced many-body electronic interactions that reconstruct the spectral function into non-dispersive quasi-particle features (Fig. 3*I*). In contrast, Bi/Bi$_2$Se$_3$ will not have this effect. We believe that the different energy position of the substrate induced Bi Dirac cone in Bi/Bi$_2$Te$_3$ versus Bi/Bi$_2$Se$_3$ is related to both strain and interface effect. The strain effect is clearly reflected in the drastic different band structures of freestanding Bi bilayer at the respective Bi$_2$Te$_3$ and Bi$_2$Se$_3$ lattice constant, as shown in Fig. 4D and Fig. 5D. In addition, with the decreasing interfacial distance between the Bi bilayer and substrate, the interface interaction splits the Bi band and further modifies the energy position of the Bi Dirac cone.

The exact origin and nature of the many-body interaction that reconstructs the linear Dirac cone spectrum is not fully clear. However, we have performed some controlled experiments to rule out the electron-phonon interaction. By controlling growth condition (28), we can tune the Fermi level of the Bi$_2$Te$_3$ substrate. The Fermi level in Fig. 1B is about 50 meV higher than that in Fig. 2*A*, which also moves the position of Dirac point. On the other hand, the phonon frequency of two samples should be in the same range. If the electron-phonon coupling were significant, we would expect a change in the quasi-particle Dirac spectra since the relative energy between electron and phonon is different in the two cases. On the contrary, from the measured ARPES spectra (Fig. 1*B* vs. Fig. 2*A*), we didn't observe any noticeable change in the quasi-particle spectra. We have also done temperature-dependent experiments at 100K and 10K which additionally showed no change of quasi-particle spectra with temperature. These experimental results suggest that the electron-phonon interaction (29) is unlikely the origin.

Therefore, we think the observed quasi-particle spectra have an electronic origin. The absence of band renormalization in Bi/Bi$_2$Se$_3$ supports the view that the electronic many-body interaction in Bi/Bi$_2$Te$_3$ is associated with the hybridization of TI states, based on our comparative DFT calculations between the two systems. Another significant difference between the two systems is strain, which may play an important role in affecting the degree of many-body interaction. But the exact form of the many-body interaction remains unclear and deserves further investigation. It can be either the Columbic electron-electron interaction or electron-plasmon interaction. For the electron-plasmon interaction, usually it shows up with satellite diamond-shaped plasmaron bands between the two Dirac points, as observed in graphene (4, 6). So far, we have not observed the diamond spectral shape but instead a vertical non-dispersive feature with comparable energy and momentum resolution with Refs. 4 and 6. However, the effects of disorder, sample quality and ARPES resolution may have prevented us from observing the plasmaron bands.

**Materials and Methods**

1) Experimental Method:

$Bi_2Te_3$ and $Bi_2Se_3$ thin films and bulk single crystals with different Fermi energy are used as substrates. $Bi_2Se_3$ films up to 40 QLs are grown by molecular beam epitaxy (MBE) method on Si (111) wafer. Bulk single crystals are grown by modified Bridgman method. Single crystals were cleaved in situ at 10K resulted in shiny flat and well-ordered surfaces. Bi films were grown on TI substrates in situ at 200K. The thickness of Bi films was monitored by reflection high-energy electron diffraction (RHEED) and scanning tunneling microscopy (STM). The sample temperature was kept at 100K and/or 10K during measurement. ARPES measurements were performed with in-lab He discharge lamp (He-I 21.2 eV), 28-90 eV photons at Advanced Light Source (ALS) beam lines 12.0.1 and ARPES beamline in National Synchrotron Radiation Laboratory (NSRL, Hefei) using Scienta R4000 analyzers with base pressures better than $5 \times 10^{-11}$ torr. Energy resolution is better than 15 meV and angular resolution is better than 0.02 Å$^{-1}$.

2) Computational Method:

DFT calculations for Bi (111) bilayer on $Bi_2Te_3$ and $Bi_2Se_3$ are carried out in the framework of the Perdew-Burke-Ernzerhof-type generalized gradient approximation using VASP package (30). The lattice parameters of the substrate were taken from experiments ($a$ =4.386 Å for $Bi_2Te_3$ and $a$ =4.138 Å for $Bi_2Se_3$), and the Bi bilayer is strained to match the substrate lattice parameter. All calculations are performed with a plane-wave cutoff of 600eV on an 11×11×1 Monkhorst-Pack k-point mesh. The substrate is modeled by a slab of 6QL $Bi_2Te_3$ and $Bi_2Se_3$, and the vacuum layers are over 20 Å thick to ensure decoupling between neighboring slabs. During structural relaxation, atoms in the lower 4QL substrate are fixed in their respective bulk positions, and the Bi bilayer and upper 2QL of substrate are allowed to relax until the forces are smaller than 0.01 eV/Å.

**Acknowledgements**

The experimental work conducted at Shanghai Jiaotong University is supported by National Basic Research Program of China (Grants No. 2012CB927401, 2011CB921902, 2011CB922200), NSFC (Grants No. 91021002, 10904090, 11174199, 11134008, 11274228), SSTCC (No. 09JC1407500, 10QA1403300, 10JC1407100, 10PJ1405700, 12JC1405300). The theoretical work conducted at University of Utah is supported by US DOE-BES (Grant No. DE-FG02-04ER46148). D.Q. acknowledges additional support from "ShuGuang" project supported by Shanghai Municipal Education Commission and Shanghai Education Development Foundation and from Program for Professor of Special Appointment (Eastern Scholar) at Shanghai Institutions of Higher Learning. Z.F.W. acknowledges additional support from ARL (Cooperative Agreement No. W911NF-12-2-0023). M.M. acknowledges additional support from NSF-MRSCE (Grant No. DMR-1121252). The Advanced Light Source is supported by the Director, Office of Science, Office of Basic Energy Sciences, of the U.S. Department of Energy under Contract No. DE-AC02-05CH11231.

**Figure legends:**

**Fig.1.** Experimental band dispersions along high symmetry directions. (*A*) ARPES spectra of 40 quintuple layers (QLs) $Bi_2Te_3$ film along K-Γ-K cut. Green lines mark the linearly dispersive "V" shape surface bands. The Dirac point "D" at the binding energy of ~ 0.2 eV is hidden by the "M" shape bulk state (BS) valence bands. The Fermi level lies in the bulk energy gap. No bulk conduction bands were observed. The inset is the first surface Brillouin zone of the system. K and M are the high symmetric points. (*B*) ARPES spectra of one bilayer Bi(111) film on $Bi_2Te_3$ along K-Γ-K and (*C*) along M-Γ-M. The "V" and "Λ" shape bands split vertically away from the original Dirac point "D" of bare $Bi_2Te_3$ film. The "M" shape BS bands disappear. (*D*) High resolution ARPES spectra showing the two split upper "V" and lower "Λ" bands, as marked by green lines and (*E*) corresponding momentum distribution curves (MDCs) showing the non-dispersive feature between the "V" and "Λ" bands. Red dots mark the band dispersions from MDCs fitting. (*F*) ARPES spectra of bare $Bi_2Se_3$. A sharp Dirac point is seen at ~ 0.3 eV below Fermi level. Due to intrinsic n-doping, bulk conduction bands were observed surrounded by surface state (SS) bands. (*G*) ARPES spectra of one bilayer Bi(111) on $Bi_2Se_3$ along K-Γ-K and (*H*) along M-Γ-M. Two Dirac points are seen as marked $D_S$ and $D_{Bi}$.

**Fig. 2.** Photon energy dependence of the ARPES spectra. (*A*)-(*C*) Bi/ $Bi_2Te_3$. Green lines mark the linearly dispersive bands and the non-dispersive feature. (*D*)-(*F*) Bi/$Bi_2Se_3$. Blue lines mark the Dirac cone from $Bi_2Se_3$; green lines mark the new Dirac cone from Bi bilayer. By changing incident photon energy, the $k_z$ of the detected energy bands changes. The observed dispersion relations of all the linearly dispersive bands don't change at all, which indicates their 2D characters. Relative intensity of the bands changes under different photon energy because of the photoemission matrix element effects and/or changing electron escape length.

**Fig. 3.** Theoretical energy bands and self-energy correction. (*A*) and (*B*) Bi/Bi$_2$Te$_3$ (only contributions from top Bi bilayer are plotted). (*C*) and (*D*) Bi/Bi$_2$Se$_3$ (the contributions from top Bi bilayer plus upper 2QL Bi$_2$Se$_3$ are plotted). (*E*) Experimental bands of Bi/Bi$_2$Te$_3$ along K-Γ-K cut superimposed with theoretical bands (Green open circles). (*F*) Experimental bands of Bi/Bi$_2$Se$_3$ along M-Γ-M superimposed with theoretical bands (Green open circles). The Fermi levels are shifted to the same position. (*G*) and (*H*) Magnified theoretically calculated Dirac cones without (*G*) and with (*H*) self-energy correction for Bi/Bi$_2$Te$_3$. (*I*) The experimental quasi-particle spectrum superimposed with theoretical spectrum extracted from (*H*) (Green open circles) to illustrate the non-dispersive feature near Dirac point. Note that the DFT bands (*A-D*, *G* and *H*) are broadened by a Lorentzian width of ~20meV for better showing the spectral functions, but they are not to be confused with self-energy correction (*H*), because they will only uniformly broaden the width of all the bands but not the relative width nor the position of the bands as the self-energy correction will do.

**Fig. 4.** Model theoretical energy bands as a function of interfacial distance for Bi/Bi$_2$Te$_3$ along M-Γ-M. (*A*)-(*D*) Contributions from top Bi bilayer and (*E*)-(*F*) Contributions from top Bi bilayer plus upper 2QL Bi$_2$Te$_3$. At the equilibrium position (*E*), the Bi bilayer has ~50% spectral weight at the hybrid Dirac point (D$_h$).

**Fig. 5.** Model theoretical energy bands as a function of interfacial distance for Bi/Bi$_2$Se$_3$ along M-Γ-M. (*A*)-(*D*) Contributions from top Bi bilayer and (*E*)-(*F*) Contributions from top Bi bilayer plus upper 2QL Bi$_2$Se$_3$. At the equilibrium position (*E*), the Bi bilayer has ~90% spectra weight at the Dirac point (D$_{Bi}$) and the 6QL Bi$_2$Se$_3$ has ~100% spectral weight at the Dirac point (D$_S$).

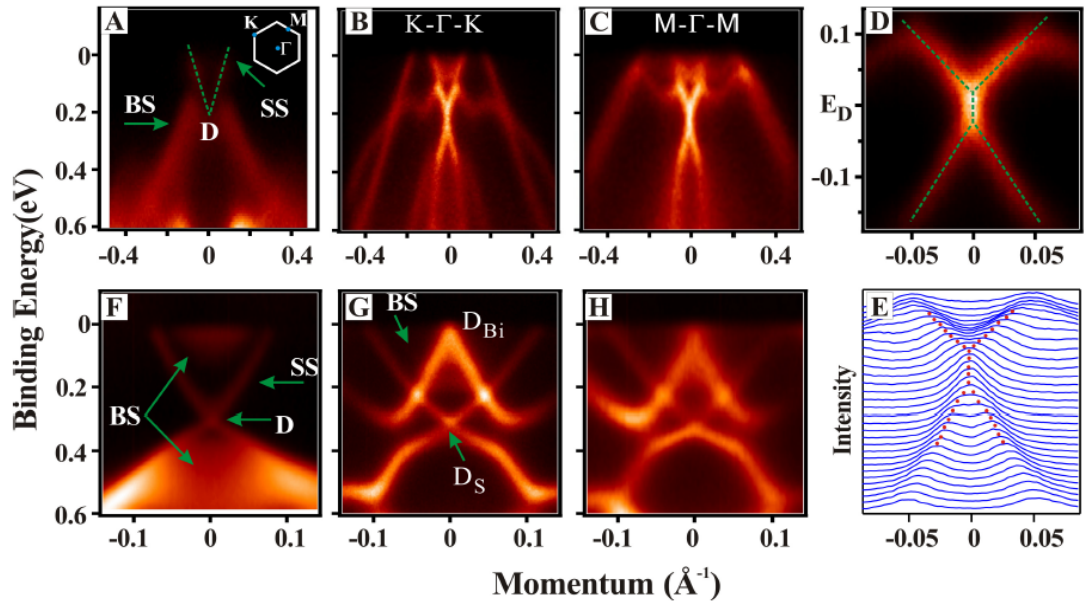

Figure 1

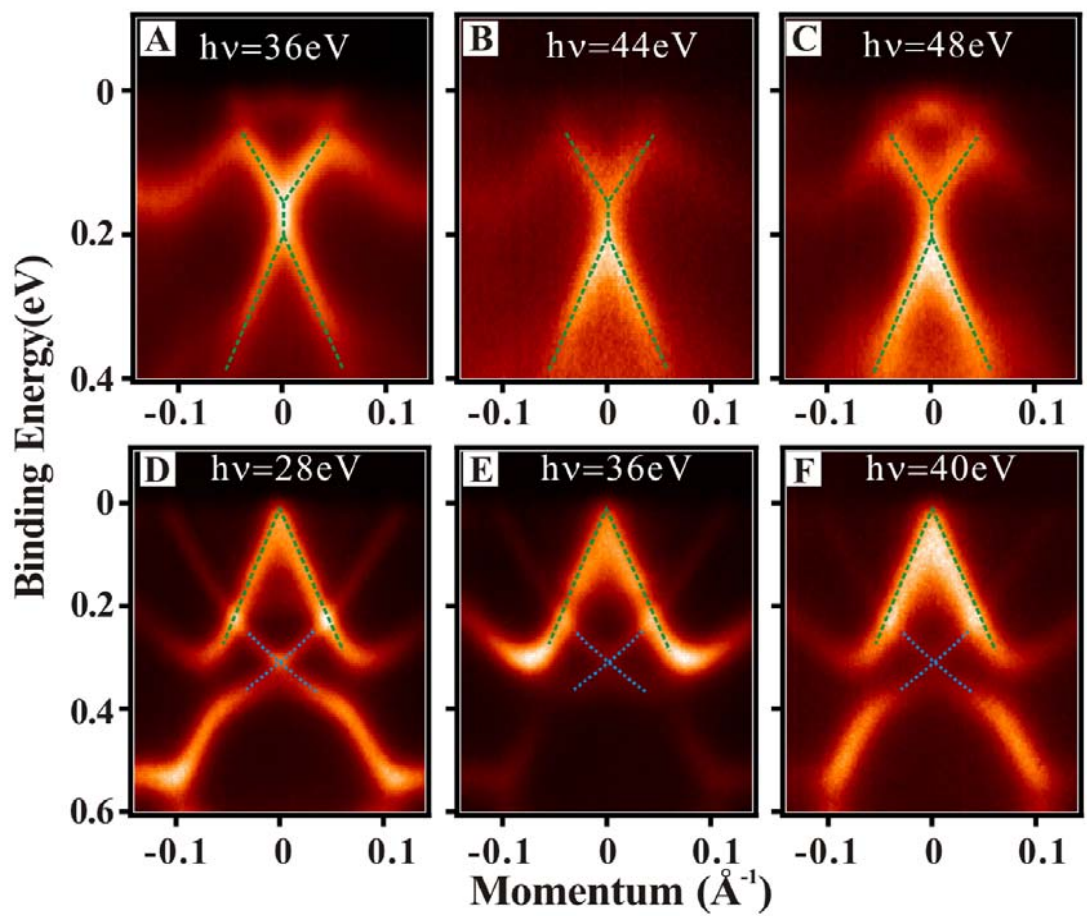

Figure 2

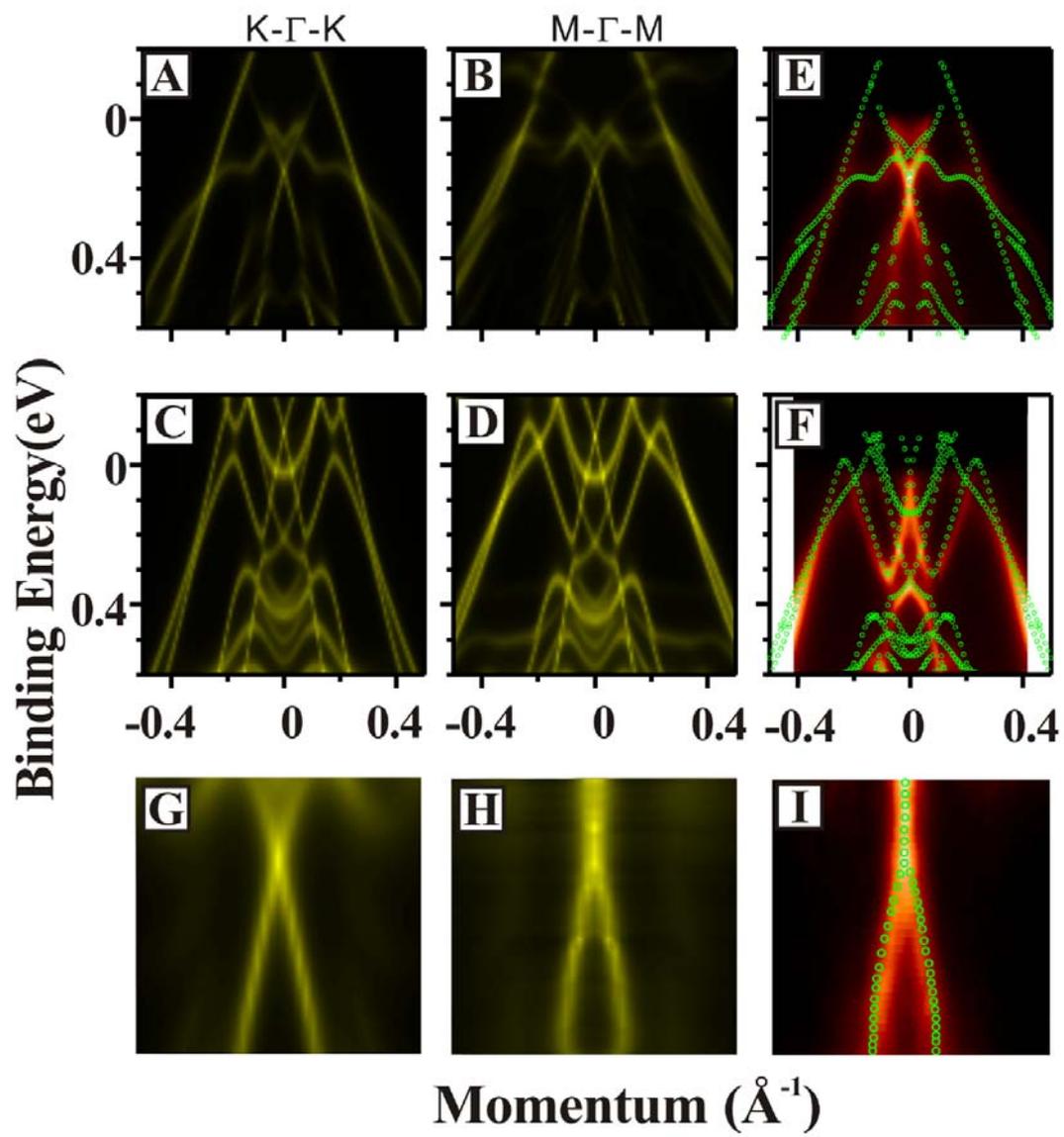

Figure 3

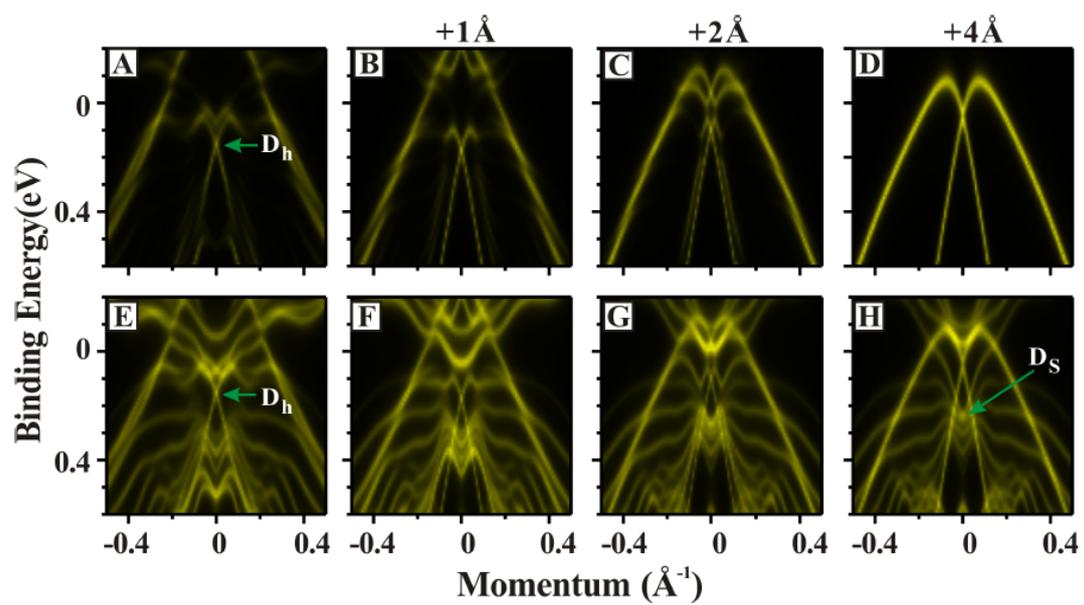

Figure 4

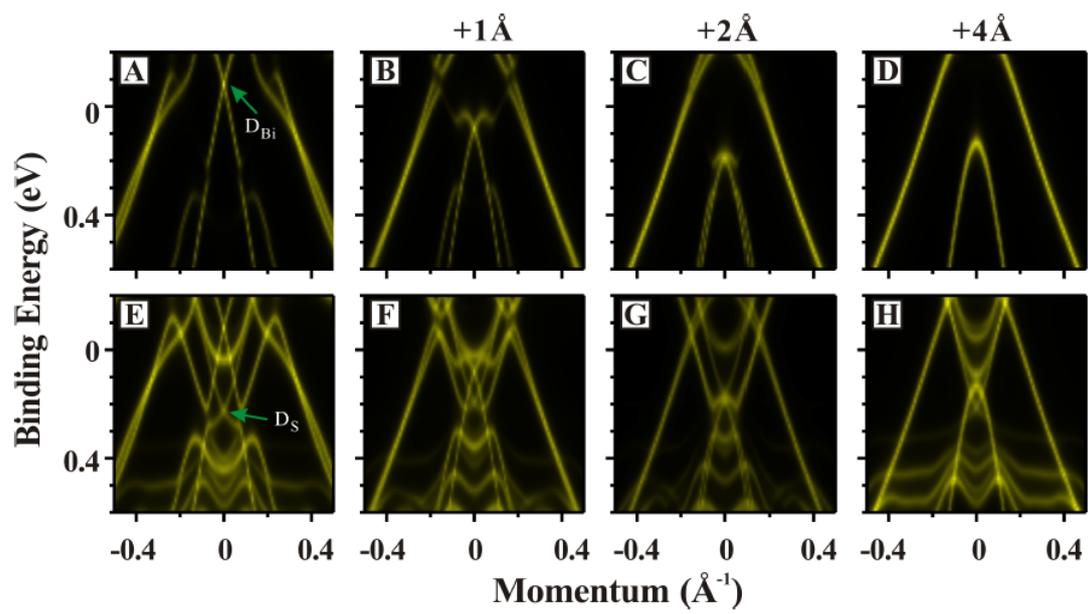

Figure 5